\documentclass[aps,pra,onecolumn,showpacs]{revtex4}
\usepackage{graphicx}
\begin{document}
\title{Number and spin densities in the ground state of a trapped mixture of two pseudospin-$\frac{1}{2}$ Bose gases with interspecies spin-exchange interaction  }
\author{Jinlong Wang}
\affiliation{Department of Physics and State Key Laboratory of Surface Physics, Fudan University, Shanghai 200433, China}
\author{Yu Shi}
\thanks{
yushi@fudan.edu.cn}
\affiliation{Department of Physics and State Key Laboratory of Surface Physics, Fudan University, Shanghai 200433, China}

\begin{abstract}
We consider  the ground state of a mixture of two pseudospin-$\frac{1}{2}$ Bose gases with interspecies spin exchange in a trapping potential. In the mean field approach, the ground state can be described in terms of four wave functions governed by a set of coupled  Gross-Pitaevskii-like (GP-like) equations, which differ from the usual GP equations in the existence of an interference term due to spin-exchange coupling between the two species.  Using these GP-like equations,  we calculate such ground state properties as chemical potentials, density profiles and spin density profiles, which are directly observable in experiments. We compare the cases with and without spin exchange.  It is demonstrated that the spin exchange  between the two species  lowers the chemical potentials, tends to equalize the wave functions of the two pseudospin components of each species,   and thus homogenizes the spin density. The novel features of the density and spin density profiles can serve as experimental probes of this novel Bose system.

\end{abstract}
\maketitle
\section{Introduction}

Multicomponent Bose-Einstein condensation (BEC) has been an active subject of research in recent years. People have considered BEC of a mixture of two different spinless species~\cite{BECD,homix,pu,ao,esry,timmermans,trippenbach,myatt} and of spinor gases such as spin-1~\cite{ho1,CKL,ueda,TLH4031,spinor} and  pseudospin-$\frac{1}{2}$ gases~\cite{ABK170403,SA063612,li}.
More recently,  a mixture of two distinct species of pseudospin-$\frac{1}{2}$ gases with interspecies spin exchange was investigated theoretically~\cite{shi3007,shi140401,shi60008,shi013637,wu1,ge1,ge2}. In such a mixture, there are $N_a$ atoms of species $a$ and $N_b$ atoms of species $b$, while each atom has a pseudospin degree of freedom  with basis state $\sigma=\uparrow,\downarrow$.  The particle number of each species $N_\alpha=N_{\alpha\uparrow}+N_{\alpha\downarrow}$, ($\alpha=a,b$), is conserved, but the particle number of each pseudospin component of each species $N_{\alpha\sigma}$, ($\sigma=\uparrow, \downarrow$),  is not conserved because of the spin-exchange coupling between the two species.
Note that for a pseudospin-$\frac{1}{2}$ gas, the total spin of each species is always a constant  $S_{\alpha}=N_{\alpha}/2$, while its $z$-component is $S_{\alpha z}=(N_{\alpha\uparrow}-N_{\alpha\downarrow})/2$.

The scattering between any two atoms are now pseudospin-dependent. For the scattering between an atom of species $\alpha$  incoming with pseudospin $\sigma_1$ and outgoing with pseudospin $\sigma_4$ and an atom of species $\beta$ incoming with pseudospin $\sigma_2$ and outgoing with pseudospin $\sigma_3$, the scattering length is  denoted $\xi^{\alpha\beta}_{\sigma_4\sigma_3\sigma_2\sigma_1}$. Correspondingly, the effective interaction is $g^{\alpha\beta}_{\sigma_4\sigma_3\sigma_2\sigma_1}\delta(\mathbf{r}-\mathbf{r}')$, where  $g^{\alpha\beta}_{\sigma_4\sigma_3\sigma_2\sigma_1} \equiv
2\pi\hbar^2\xi^{\alpha\beta}_{\sigma_4\sigma_3\sigma_2\sigma_1}/\mu_{\alpha\beta}$, where $\mu_{\alpha\beta}$ is the reduced mass of the two atoms. For convenience, we define the  shorthands
$g^{\alpha\alpha}_{\sigma\sigma }\equiv g^{\alpha\alpha}_{\sigma\sigma\sigma\sigma}$  for the intraspecies scattering of the same pseudospin $\sigma$, $g^{\alpha\alpha}_{\sigma\bar{\sigma} }\equiv
2g^{\alpha\alpha}_{\sigma\bar{\sigma}\bar{\sigma}\sigma}$  for the intraspecies scattering of different pseudospins  $\sigma \neq \bar{\sigma}$,  $g^{ab}_{\sigma\sigma'}\equiv g^{ab}_{\sigma\sigma'\sigma'\sigma}$ for the  interspecies scattering without spin exchange,  $g_e \equiv g^{ab}_{\sigma\bar{\sigma}\sigma\bar{\sigma}}$ for the interspecies spin-exchange scattering. We shall also use the shorthands $\xi^{\alpha\alpha}_{\sigma\sigma}$ for $\xi^{\alpha\alpha}_{\sigma\sigma\sigma\sigma}$, $\xi^{\alpha\alpha}_{\sigma\bar{\sigma}} $  for $2\xi^{\alpha\alpha}_{\sigma\bar{\sigma}\sigma\bar{\sigma}}$, $\xi^{ab}_{\sigma\sigma'}$ for  $\xi^{ab}_{\sigma\sigma'\sigma'\sigma}$. In this paper, all these scattering lengths $\xi$'s and thus the effective interaction strengths $g$'s are considered to be positive. Thus the many-body Hamiltonian density is~\cite{shi3007,shi140401}
\begin{widetext}
\begin{equation}
\begin{array}{rl}
\displaystyle
\hat{\mathcal{H}}(\mathbf{r})=&
\displaystyle  \sum_{\alpha\sigma} \hat{\psi}_{\alpha\sigma}^\dagger [-\frac{1}{2m_\alpha}\nabla^2 +   U_{\alpha\sigma}(\mathbf{r})] \hat{\psi}_{\alpha\sigma} + \frac{1}{2}\sum_{\alpha\sigma\sigma^{'}}{
g^{\alpha\alpha}_{\sigma\sigma^{'}}
|\hat{\psi}_{\alpha\sigma}|^{2}|\hat{\psi}_{\alpha\sigma^{'}}|^{2}} \\
\displaystyle
&
\displaystyle +\sum_{\sigma\sigma^{'}}{g^{ab}_{\sigma\sigma^{'}}|\hat{\psi}_{a\sigma}|^{2}
|\hat{\psi}_{b\sigma^{'}}|^{2}}+
g_{e}(\hat{\psi}_{a\uparrow}^{\dagger}
\hat{\psi}_{b\downarrow}^{\dagger}\hat{\psi}_{b\uparrow}\hat{\psi}_{a\downarrow}+
\hat{\psi}_{a\downarrow}^{\dagger}
\hat{\psi}_{b\uparrow}^{\dagger}\hat{\psi}_{b\downarrow}\hat{\psi}_{a\uparrow}),
\end{array}
\end{equation}
\end{widetext}
where $U_{\alpha\sigma}(\mathbf{r})$ is the external trapping potential,  $\hat{\psi}_{\alpha\sigma} \equiv \hat{\psi}_{\alpha\sigma}(\mathbf{r})$ is the Bose field operator for species $\alpha$ with pseudospin $\sigma$ ($\alpha = a, b$, $ \sigma =\uparrow, \downarrow$), $\sigma$ and $\sigma'$ may or may not be equal. Expanded in terms of an orthonormal set of single-particle orbital basis states $\{\phi_{\alpha\sigma,i}(\mathbf{r})\}$, \begin{equation}
\hat{\psi}_{\alpha\sigma}(\mathbf{r})=\sum_i \hat{a}_{\alpha\sigma,i}\phi_{\alpha\sigma,i}(\mathbf{r}), \label{expansion}
\end{equation}
where   $\hat{a}_{\alpha\sigma,i}$ is the annihilation operator corresponding to $\phi_{\alpha\sigma,i}(\mathbf{r})$.  In  $\hat{\mathcal{H}}$, the first two summations of  interaction terms are the density-density interactions  without spin exchange, as studied in previous models of Bose mixtures, the last term is the spin-exchange interaction, which causes spin correlation or entanglement between the two species, and is also responsible for the novel features discussed in the present paper.

Under the usual single orbital-mode approximation, Bose statistics and energetics governs that all atoms of each species $\alpha$ and with the pseudospin state $\sigma$ occupy the lowest-energy single-particle orbital mode hereby denoted as $\phi_{\alpha\sigma}(\mathbf{r})$, hence in the expansion (\ref{expansion}) of the field operator $\hat{\psi}_{\alpha\sigma}(\mathbf{r})$, one only needs to consider one term $\hat{a}_{\alpha\sigma}\phi_{\alpha\sigma}(\mathbf{r})$, where $\hat{a}_{\alpha\sigma}$ is the annihilation operator corresponding to    $\phi_{\alpha\sigma}(\mathbf{r})$. Consequently, in each term of the many-body Hamiltonian $\int d^3r \hat{\mathcal{H}}(\mathbf{r})$, there is an integration of a product of single-particle wave functions, which now becomes an effective coefficient. Therefore, under the single orbital-mode approximation, the details of $\phi_{\alpha\sigma}(\mathbf{r})$ are not needed  in describing the many-body  ground state  in terms of creation and annihilation operators or, equivalently, the collective spin operators, although such simplification is lost when one goes beyond single orbital-mode approximation.  In a broad parameter regime, the two species are quantum entangled in the particle numbers of the two pseudospin states or  collective spins, that is, the two species do not undergo BEC separately, hence the ground state was dubbed entangled BEC.

However, these four wave functions and the corresponding elementary excitations  are important physical properties.  For a uniform system,  $\phi_{\alpha\sigma}(\mathbf{r})$  is simply the constant $1/\sqrt{\Omega}$, where $\Omega$ is the volume of the system.   For convenience, we write the mean field value of the Bose field operator  $\langle \hat{\psi}_{\alpha\sigma}\rangle$ as $\psi_{\alpha\sigma}e^{i\gamma_{\alpha\sigma}}$, where $\psi_{\alpha\sigma} >0$. This is the so-called condensate wave function. The condensate wave function for pseudospin $\sigma$ component of species $\alpha$ is $\psi_{\alpha\sigma} = \eta_{\alpha\sigma} \sqrt{N_{\alpha\sigma}^0}/\sqrt{\Omega}$, where $N_{\alpha\sigma}^0$ is the corresponding particle number in the many-body ground state of the system, determined by the many-body Hamiltonian, $\eta_{\alpha\sigma} = \pm $ is a sign. To minimize the energy, the signs of three components can be chosen to be $+$ while that of  the other one is chosen to be $-$.

In a trapping potential, which is an experimental necessity for BEC of cold atoms, the wave function $\psi_{\alpha\sigma}(\mathbf{r})$    is dramatically different from a constant. Moreover, the details of  $\psi_{\alpha\sigma}(\mathbf{r})$ provide experimentally very important information, as their modular square is just the particle density, which is directly measurable and is a key  observable.

In this paper,  we consider such a pseudospin-$\frac{1}{2}$ mixture in a trapping potential, and find some interesting properties, especially the density and spin density profiles of the four lowest-energy orbital modes $\{ \phi_{\alpha\sigma} \}$. There had been many calculations on such properties in other types of  BEC mixtures~\cite{pu,Edwards}, which demonstrated that a trapping potential brings significant features absent in a homogeneous system.

The GP-like equations can be obtained by  using the Euler-Lagrange equation
\begin{eqnarray}
&&\frac{d}{dt}\left(\frac{\partial\mathcal{L}}{\partial\dot{\psi}_{\alpha\sigma}}\right)
-\frac{\partial\mathcal{L}}{\partial\psi_{\alpha\sigma}}=0,
\end{eqnarray}
with $\mathcal{L}=
i \sum_{\alpha\sigma}\psi_{\alpha\sigma}^* \partial_t \psi_{\alpha\sigma}
-\langle \hat{\mathcal{H}} \rangle $, and then substituting $i\partial_t$ as $\mu_{\alpha\sigma}$.  One obtains
\begin{widetext}
\begin{eqnarray}
&&\left(-\frac{\hbar^2}{2m_\alpha}\nabla^2+U_{\alpha\sigma}(\mathbf{r})\right)\psi_{\alpha\sigma}
   +g_{\sigma\sigma}^{\alpha\alpha}|\psi_{\alpha\sigma}|^2\psi_{\alpha\sigma}+ g_{\sigma\bar{\sigma}}^{\alpha\alpha}|\psi_{\alpha\bar{\sigma}}|^2\psi_{\alpha\sigma}+
   g_{\sigma\sigma}^{\alpha\bar{\alpha}}|\psi_{\bar{\alpha}\sigma}|^2\psi_{\alpha\sigma}\nonumber\\
&&\qquad+ g_{\sigma\bar{\sigma}}^{\alpha\bar{\alpha}}|\psi_{\bar{\alpha}\bar{\sigma}}|^2\psi_{\alpha\sigma}
   - g_e\psi_{\bar{\alpha}\bar{\sigma}}^*
   \psi_{\bar{\alpha}\sigma}\psi_{\alpha\bar{\sigma}}
   =\mu_{\alpha\sigma}\psi_{\alpha\sigma}. \label{GPphi}
\end{eqnarray}
\end{widetext}
where $g_e$ term is due to the spin-exchange interaction, and is a new feature absent in previous models of Bose mixtures.  The minus sign comes from the requirement that the phases $\gamma_{\alpha\sigma}$ of the four components should satisfy $\cos(\gamma_{a\uparrow}-\gamma_{a\downarrow}-\gamma_{b\uparrow}
+\gamma_{b\downarrow})=-1$ for the  minimization of the energy.  For $g_e >0$,  this spin-exchange interaction is like an attractive interaction in some way, counteracting the other interaction terms if the latter is repulsive. However,  it is a new effect, as it depends on the wave functions rather than the densities.

When $g_e$ term is negligible, the system behaves like the usual Bose mixtures with repulsive interactions. When $g_e$ term is dominant, the system behaves in a way similar to an attractive mixture with intraspecies interaction negligible. Moreover, to minimize the spin-exchange interaction energy
$g_{e}\langle \hat\psi_{a\uparrow}^{\dagger}\hat\psi_{b\downarrow}^{\dagger}\hat\psi_{b\uparrow}
\hat\psi_{a\downarrow}+
\hat\psi_{a\downarrow}^{\dagger}
\hat\psi_{b\uparrow}^{\dagger}\hat\psi_{b\downarrow}\hat\psi_{a\uparrow}\rangle=-2g_e \psi_{a\uparrow}
\psi_{a\downarrow}\psi_{b\downarrow}\psi_{b\uparrow} \cos(\gamma_{a\uparrow}-\gamma_{a\downarrow}-\gamma_{b\uparrow}
+\gamma_{b\downarrow})=-2g_e \psi_{a\uparrow}
\psi_{a\downarrow}\psi_{b\downarrow}\psi_{b\uparrow}  $, $\psi_{a\uparrow}
\psi_{a\downarrow}\psi_{b\downarrow}\psi_{b\uparrow}$ should be maximized, as a kind of interference effect, which means that for each species $\alpha=a,b$,  $\psi_{\alpha\uparrow}(\mathbf{r})=\psi_{\alpha\downarrow}(\mathbf{r})$, hence the density profiles for the two pseudospin components of each species tends to be the same. To see this, note that each wave function can be taken to be real and positive~\cite{feynman}. The other terms in the Hamiltonian of course break this equality, as will be studied later in this paper.

Below we shall describe the finding that the larger the interspecies spin exchange $g_e$ is, the stronger the overlap between the  density profiles of the two pseudospin components of each species is. Hence density profiles are very good experimental probes of the underlying interspecies correlations. On the other hand, by comparing the experimental and theoretical results on the number density and spin density profiles, one may estimate the spin-exchange interaction strength $g_e$. Experimentally, by studying the the effect of $g_e$ on the density profiles, one can obtain the information on such a mixture.

Up to now, there is not yet a report on experimental studies of such a spin-exchange mixture between different species.  However, the interspecies spin-exchange interaction is determined by the difference between the interspecies triplet and singlet scattering lengths, which has been found to be quite a few nanometer (nm)~\cite{ferrari}. In this paper, the theoretical investigation using this parameter value  clearly indicates interesting new features. Hence our work also provides some motivation and methodology for experimental exploration of such a mixture.

In Sec.~\ref{nume}, the numerical method is described.  In Sec.~\ref{calculations}, we calculate the ground state properties, comparing the cases with and without spin exchange, and demonstrating the experimentally observable effects of interspecies spin exchange. Then we make a summary in Sec.~\ref{summary}.

\section{Numerical method  \label{nume}}

We assume  the trapping potential to be
\begin{eqnarray}
&&U_{\alpha\sigma}(\mathbf{r})=\frac{1}{2}M_\alpha\omega_\alpha^2(\rho^2+\lambda^2 z^2), \label{harmo} \label{potential}
\end{eqnarray}
where $\rho=\sqrt{x^2+y^2}$, $\lambda$ represents the trap anisotropy, $M_\alpha$ and  $\omega_{\alpha}$ are the mass of $\alpha$-atom and trap frequency, respectively.   $U_{\alpha\uparrow}=U_{\alpha\downarrow}$. For a magnetic trap, $M_\alpha\omega_\alpha^2=\gamma_{\alpha}\mu_B B_0$, with  $\gamma_{\alpha}$ being the $g$ factor of $\alpha$-atom, $B_0$ being the  central magnetic field multiplied by a normalized factor.
In order that the parameter values are close to the experimental data, we imagine species $a$ as ${}^{87}$Rb and species
$b$   as ${}^{23}$Na, then $M_a\omega_a^2/M_b\omega_b^2=\gamma_{a}/\gamma_{b}=1$, {\rm i.e.} $U_{a\sigma}=U_{b\sigma}$. Define   $\kappa\equiv \omega_b/\omega_a \sqrt{M_a/M_b} = \sqrt{87/23}$.      In our calculation, we use the parameter values $\omega_a=2\pi\times 75$Hz, $\omega_b= \kappa \omega_a$ and $\lambda=\sqrt{8}$.  The values of scattering lengths are set to be
$\xi_{\uparrow\uparrow}^{aa}=8$ nm, $\xi_{\downarrow\downarrow}^{aa}=6.7$ nm, $\xi_{\uparrow\downarrow}^{aa}=3.7$ nm, $\xi_{\uparrow\uparrow}^{bb}=4$ nm, $\xi_{\downarrow\downarrow}^{bb}=2$ nm, $\xi_{\uparrow\downarrow}^{bb}=1.7$ nm,
$\xi_{\uparrow\uparrow}^{ab}=1.2$ nm, $\xi_{\downarrow\downarrow}^{ab}=0.67$ nm, $\xi_{\uparrow\downarrow}^{ab}=1.9$ nm. We vary the value of $\xi_e$ and study its effect on the number densities and the spin densities.

We expand $\psi_{\alpha\sigma}$ in terms of $N_{basis}$ eigenfunctions of the  non-interacting Schr\"{o}dinger equation in an anisotropic harmonic potential (\ref{potential}), that is,
\begin{eqnarray}
&&   \psi_{\alpha\sigma}(\mathbf{r})= \sum_{r=0}^{N_{basis}}A^r_{\alpha\sigma} R_{n_r}^{m_r}(\rho) \Phi_{m_r}(\varphi) Z_{w_r}(z),\label{expan}
\end{eqnarray}
where $R_{n_r}^{m_r}(\rho)$, $\Phi_{m_r}(\varphi)$ and $Z_{w_r}(z)$  correspond to the cylindrical coordinates  $\rho$, $\varphi$ and $z$, respectively, $A^r_{\alpha\sigma}$ is the expanding coefficient under the condition $\sum^{N_{basis}}_r(A^r_{\alpha\sigma})^2=N_{\alpha\sigma}$. For the ground state, only eigenfunctions with $m_r=0$ are relevant.

Therefore, GP-like equation (\ref{GPphi}) is  transformed to the following nonlinear equation
\begin{widetext}
\begin{eqnarray}
&&(E^l_{\alpha\sigma}-\mu_{\alpha\sigma})A^l_{\alpha\sigma}+g_{\sigma\sigma}^{\alpha\alpha}\sum_{ijk}A^i_{\alpha\sigma}A^j_{\alpha\sigma}A^k_{\alpha\sigma} I(i\alpha,j\alpha,k\alpha,l\alpha)\nonumber\\
&&+g_{\sigma\bar{\sigma}}^{\alpha\alpha}\sum_{ijk}A^i_{\alpha\bar{\sigma}}A^j_{\alpha\bar{\sigma}}A^k_{\alpha\sigma} I(i\alpha,j\alpha,k\alpha,l\alpha)+g_{\sigma\sigma}^{\alpha\bar{\alpha}}\sum_{ijk}A^i_{\bar{\alpha}\sigma}A^j_{\bar{\alpha}\sigma}A^k_{\alpha\sigma} I(i\bar{\alpha},j\bar{\alpha},k\alpha,l\alpha)\nonumber\\
&&+g_{\sigma\bar{\sigma}}^{\alpha\bar{\alpha}}\sum_{ijk}A^i_{\bar{\alpha}\bar{\sigma}}A^j_{\bar{\alpha}\bar{\sigma}}A^k_{\alpha\sigma}I(i\bar{\alpha},j\bar{\alpha},k\alpha,l\alpha)
-g_e\sum_{ijk}A^i_{\bar{\alpha}\bar{\sigma}}A^j_{\bar{\alpha}\sigma}A^k_{\alpha\bar{\sigma}}I(i\bar{\alpha},j\bar{\alpha},k\bar{\alpha},l\bar{\alpha})=0,\label{GPEN}
\end{eqnarray}
where $E^l_{\alpha\sigma}=((2n_l+m_l+1)+(w_l+1/2)\lambda)\hbar\omega_\alpha$,

\begin{eqnarray}
&&I(i\alpha,j\alpha,k\beta,l\beta)\equiv \frac{1}{2\pi}\int_0^\infty R_{n_i}^{m_i}(\eta_{\rho \alpha}^2\rho^2)R_{n_j}^{m_j}(\eta_{\rho \alpha}^2\rho^2)R_{n_k}^{m_k}(\eta_{\rho \beta}^2\rho^2)R_{n_l}^{m_l}(\eta_{\rho \beta}^2\rho^2)\rho d\rho\nonumber\\
&&\times\int_{-\infty}^{+\infty}Z_{w_i}(\eta_{z\alpha}z)Z_{w_j}(\eta_{z\alpha}z)
Z_{w_k}(\eta_{z\beta}z)Z_{w_l}(\eta_{z\beta}z)dz, \nonumber
\end{eqnarray}
\end{widetext}
where $m_r$ ($r=i,j,k,l$) should be 0. There are various algorithms to determine the solutions of the nonlinear equation set, such as fixed-point iteration and Newton method.
We use Broyden method to obtain the solutions of Eq. (\ref{GPEN}) because of its high speed and precision.

\section{Calculations \label{calculations}}

We shall use the GP-like equations (\ref{GPphi}) to study the effect of interspecies spin exchange on the number densities $$n_{\alpha\sigma}  = |\psi_{\alpha\sigma}(\mathbf{r})|^2,$$
and the spin densities
$${\cal S}_{\alpha z}(\mathbf{r}) = \frac{1}{2} [ |\psi_{\alpha\uparrow}(\mathbf{r})|^2 - |\psi_{\alpha\downarrow}(\mathbf{r})|^2]. $$

For simplicity, we assume that the atom numbers of the two species are equal, i.e. $N_a=N_b=N$. Although the cases of $N \leq 1000$ may not be experimentally realistic, they serve as theoretical  demonstration of the dilute limit, in which the interaction energy is small, especially in comparison with the cases with larger $N$.

\subsection{The case without interspecies spin exchange}

\begin{figure*}
\scalebox{0.4}[0.4]{\includegraphics{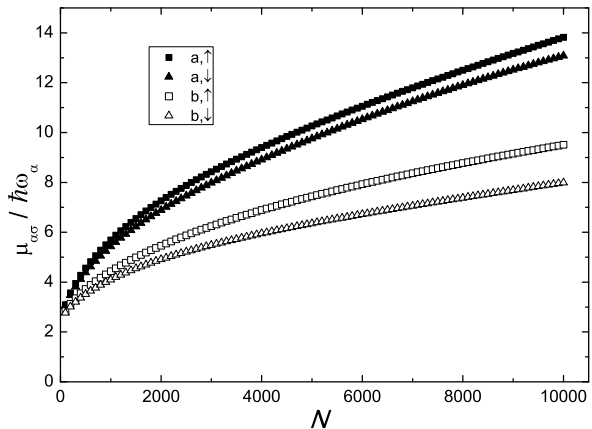}}
\caption{Reduced chemical potential $\mu_{\alpha\sigma}/ \hbar\omega_{\alpha}$ varying with the atom number $N$ of each species, at a generic parameter point without interspecies spin exchange, i.e. $\xi_e=0$.  $\mu_{\alpha\sigma}$ as well as $\mu_{a\sigma}/\hbar\omega_a - \mu_{b\sigma}/\hbar\omega_b$  increase with  $N$. } \label{uN0}
\end{figure*}

\begin{figure*}
\scalebox{0.75}[0.75]{\includegraphics{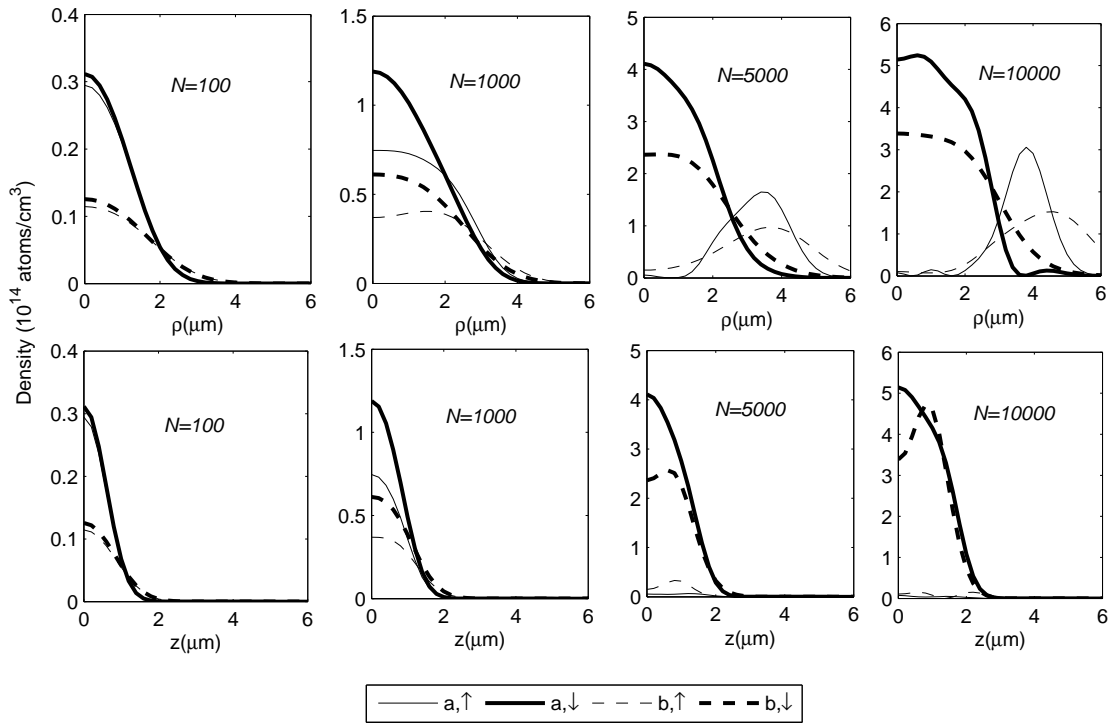}}
\caption{Density  $n_{\alpha\sigma}=|\psi_{\alpha\sigma}|^2$ for each pseudospin component of each species, at a generic parameter point with $\xi_e=0$, as defined in the text, for  $N =100, 1000, 5000, 10000$. $N$ is atom number of each species.  The upper plots are profiles along $\rho$ direction with $z=0$. The lower plots are profiles along $z$ direction with $\rho=0$.  More than one peak appear in some plots, due to the inclusion of higher order eigenfunctions in the expansion~(\ref{expan}) when the interaction energy becomes important.}\label{denrz}
\end{figure*}

\begin{figure*}
\scalebox{0.75}[0.75]{\includegraphics{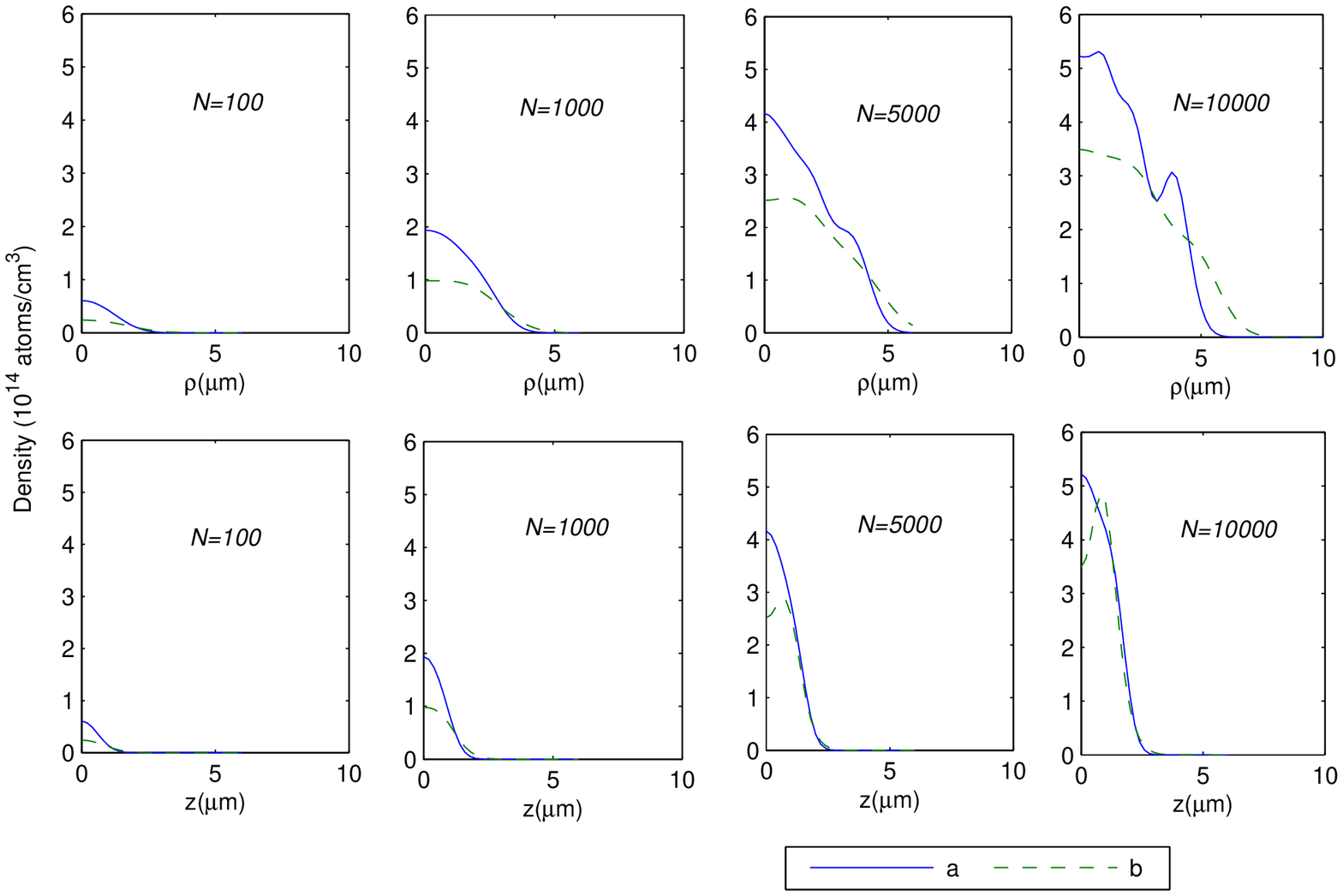}}
\caption{Profiles of the total density $n_{\alpha}=n_{\alpha\uparrow}+ n_{\alpha\downarrow}$, for each species, along $\rho$ direction with $z=0$, and along $z$ direction with $\rho=0$, at a generic parameter point with $\xi_e=0$, as defined in the text, for $N=100, 1000, 5000, 10000$.  Not all components co-exist in every region,  in order to minimize the total energy under the given parameter values.
The density profiles are more extended in  $\rho$ direction than in $z$ direction, as the trapping in $z$ direction is stronger than  in  $\rho$ direction.
 }\label{denrz01}
\end{figure*}

\begin{figure*}
\scalebox{0.7}[0.7]{\includegraphics{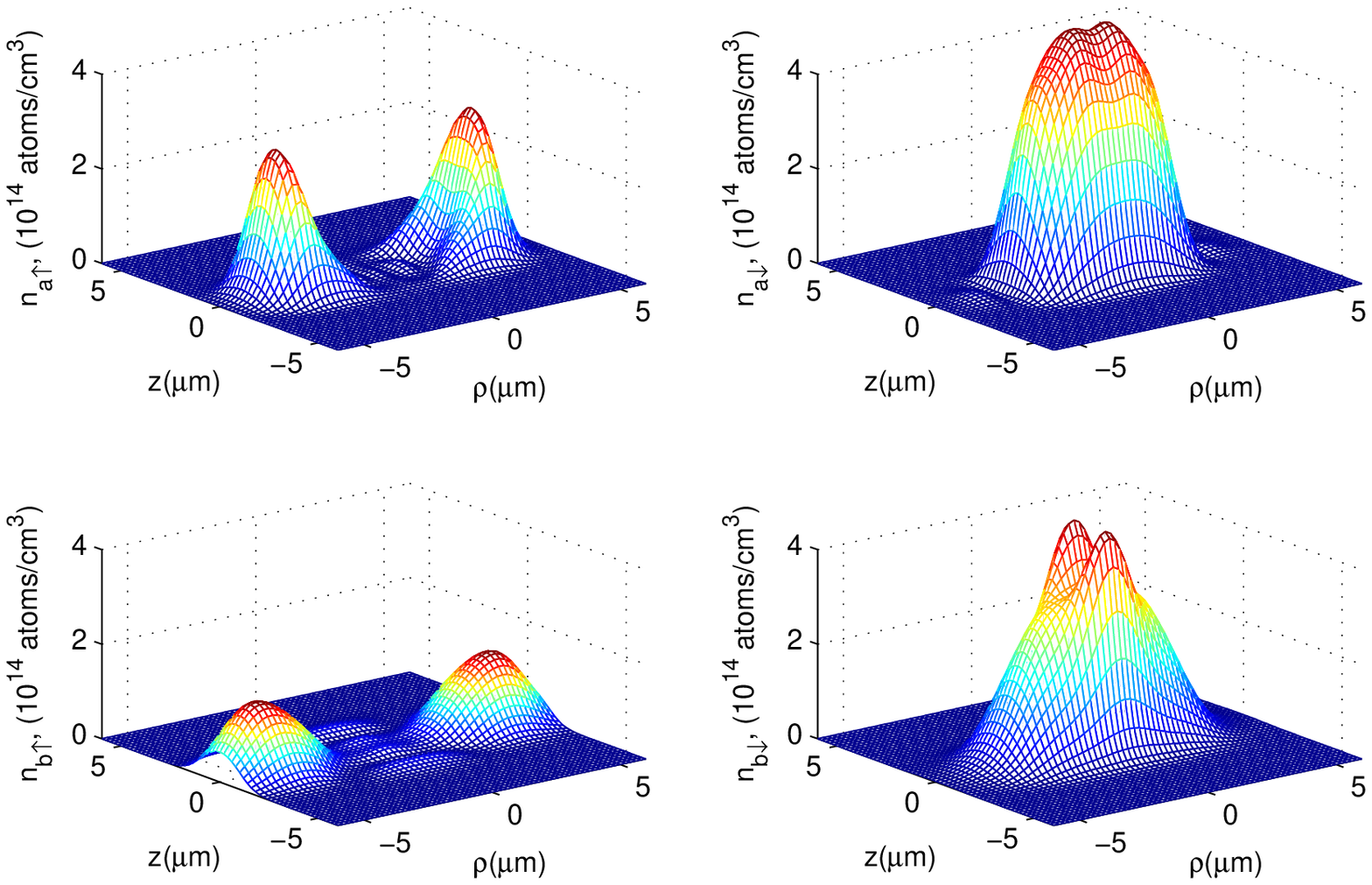}}
\caption{Two-dimensional density profiles  on a cross section including $z$-axis, with $\rho$ and $z$ as the coordinates, in a generic parameter point defined in the text, with $\xi_e=0$, $\lambda=\sqrt{8}$ and  $N=10000$.  The profiles are asymmetric between  $\rho$ and $z$ directions as $\lambda \neq 1$. } \label{den3D}
\end{figure*}

\begin{figure*}
\scalebox{0.7}[0.7]{\includegraphics{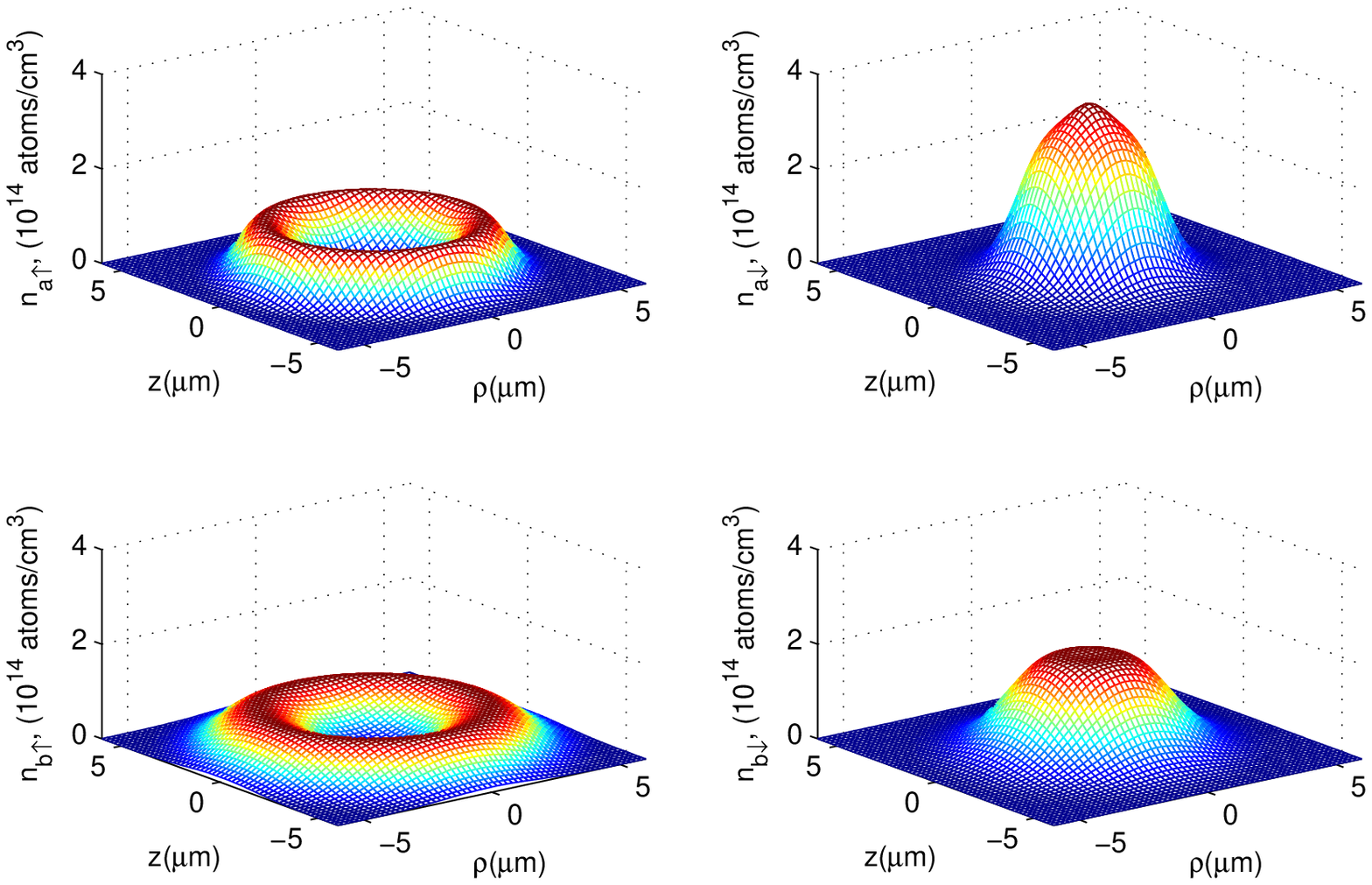}}
\caption{Two-dimensional density profiles  on a cross section including $z$-axis, with $\rho$ and $z$ as the coordinates, in a generic parameter point defined in the text, with $\xi_e=0$, $\lambda=1$ and  $N=10000$. The profiles  are symmetric  between  $\rho$ and $z$ directions   as $\lambda=1$.
 } \label{den3Dl1}
\end{figure*}

\begin{figure*}
\scalebox{0.75}[0.75]{\includegraphics{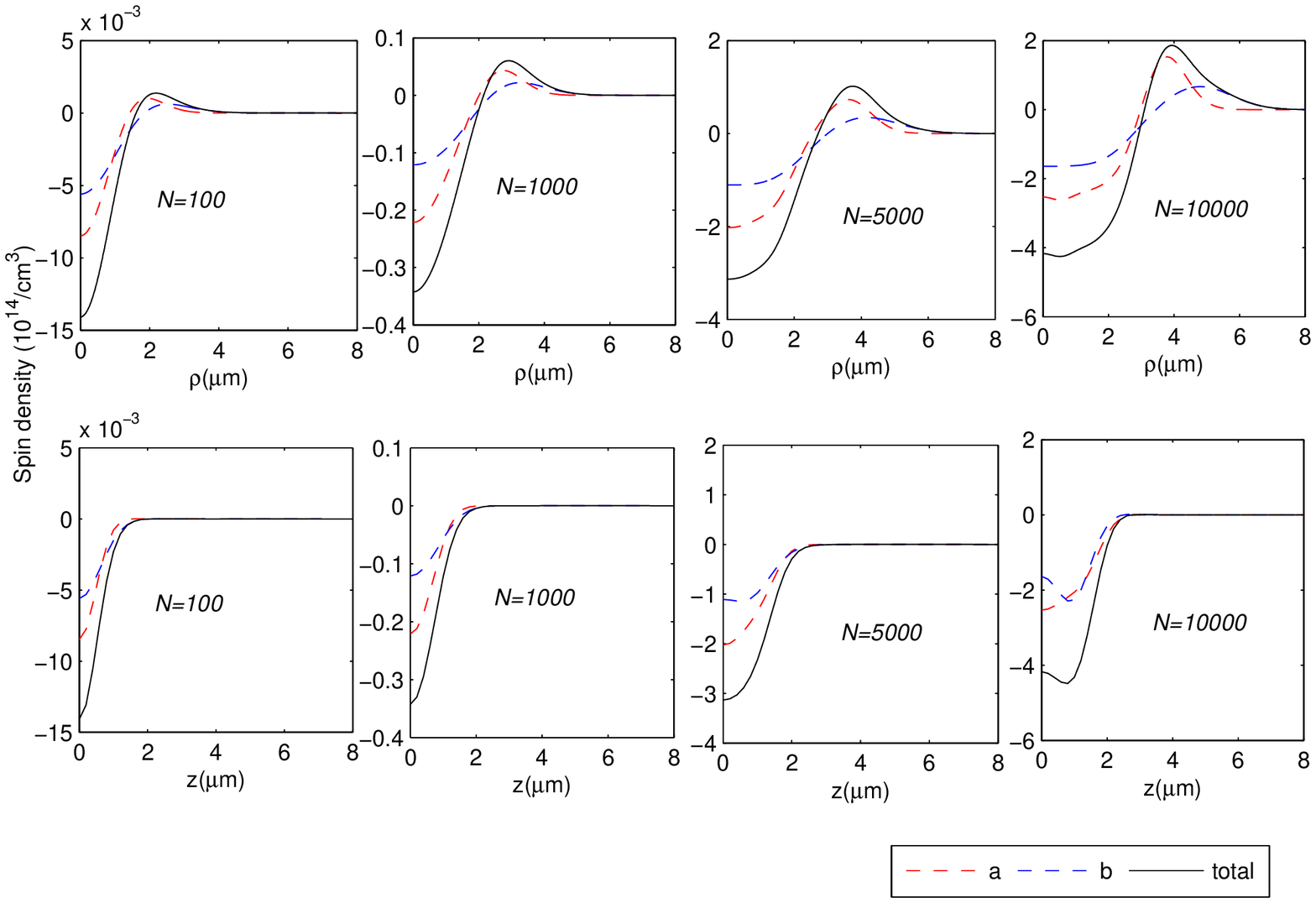}}
\caption{Spin density profiles for  $N=100, 1000, 5000, 10000$, in a generic parameter point as defined in the main text, with $\xi_e=0$. The spin density approaches zero for large enough $\rho$ or $z$, as the densities of $\uparrow$ and $\downarrow$ atoms both approach zero when the trapping potential is large enough. } \label{spinrz}
\end{figure*}

First we consider the case of $\xi_e=0$,  in which the system reduces to a mixture of four spinless condensates. The results for this case are summarized in Figures~\ref{uN0}, \ref{denrz}, \ref{denrz01}, \ref{den3D}, \ref{den3Dl1}, \ref{spinrz}.  They are all  consequences of minimizing the repulsive interactions among the four components, as detailed in the following.

We have $\mu_{\alpha\uparrow}>\mu_{\alpha\downarrow}$ under the present parameter values,
as indicated in
Fig.~\ref{uN0}, which shows the chemical potential $\mu_{\alpha\sigma}$  as a function of  $N$. $\mu_{\alpha\sigma}$ as well as $\mu_{a\sigma}/\hbar\omega_a - \mu_{b\sigma}/\hbar\omega_b$  increase with  $N$. The difference between  $\mu_{\alpha\uparrow}$ and $\mu_{\alpha\downarrow}$ is due to spin dependence of various  scattering lengths. This result can be confirmed by a calculation based on the Thomas-Fermi approximation, which leads to
$ \mu_{\alpha\sigma}=\frac{1}{\Omega}[\int U_{\alpha\sigma}d\mathbf{r}+g^{\alpha\alpha}_{\sigma\sigma} N_{\alpha\sigma} +g^{\alpha\alpha}_{\sigma\bar{\sigma}} N_{\alpha\bar{\sigma}} +g^{\alpha\bar{\alpha}}_{\sigma\sigma} N_{\bar{\alpha}\sigma} +g^{\alpha\bar{\alpha}}_{\sigma\bar{\sigma}}N_{\bar{\alpha}
\bar{\sigma}} ].$
In a sense, $\mu_{\alpha\sigma}$ represents the average energy of one atom of species $\alpha$ with pseudospin $\sigma$. We can observe that $\mu_{\alpha\sigma}/\hbar\omega_\alpha$ decreases with $N$, towards  the single particle value $1+\lambda/2=2.414$.

Now we take a look at the spatial dependence  of the atom densities $n_{\alpha\sigma}(\mathbf{r})=|\psi_{\alpha\sigma}(\mathbf{r})|^2.$
The density profiles for several different values of $N$ are  shown in
Fig.~\ref{denrz}, where the distributions along $\rho$ and $z$ directions are depicted respectively. These plots display some interesting features.   Obviously $n_{\alpha\uparrow}$ and $n_{\alpha\downarrow}$ are complementary, because of the normalization condition $\int(n_{\alpha\uparrow}+n_{\alpha\downarrow})d\mathbf{r}=N$. When $N$ is small, $n_{\alpha\uparrow}$ and $n_{\alpha\downarrow}$ are close to each other, because the interaction energy is small. But with the increase of $N$, the difference between $n_{\alpha\uparrow}$ and $n_{\alpha\downarrow}$ increases in order to lower the interaction energy. When $N$ is large enough, two or more peaks may appear in some density profiles, due to the inclusion of higher order eigenfunctions in the expansion~(\ref{expan}) when the  interaction energy becomes important.

The profiles of the total density of each species, $n_{\alpha}=n_{\alpha\uparrow}+{n_{\alpha\downarrow}}$, are shown in Fig.~\ref{denrz01}. We can see that not all components co-exist in every region, in analogy with the two-component BEC~\cite{pu}. This is in order to minimize the total energy under the given parameter values.

The density profiles are more extended in  $\rho$ direction than in $z$ direction, as exhibited in Fig.~\ref{denrz}  and in  Fig.~\ref{denrz01}.  The reason is that  the trapping in $z$ direction is stronger than that in $\rho$ direction.

We have also plotted the two-dimensional density profiles with  $\rho$ and $z$ as the two coordinates, as shown in Fig.~\ref{den3D} for $\lambda=\sqrt{8}$,  and in Fig.~\ref{den3Dl1} for $\lambda=1$, both with $N=10000$. The complementarity between the two pseudospin components of each species is very clear. In the former case, as $\lambda \neq 1$,  the profiles are asymmetric between  $\rho$ and $z$ directions. In the latter case, as $\lambda=1$, the profiles  are symmetric  between  $\rho$ and $z$ directions.

Now we consider the spin density
${\cal S}_{\alpha z}(\mathbf{r}) = \frac{1}{2} [ |\psi_{\alpha\uparrow}(\mathbf{r})|^2
-|\psi_{\alpha\downarrow}(\mathbf{r})|^2]. $
The spin density profiles are shown in Fig.~\ref{spinrz}  for $N=100, 1000, 5000, 10000$. It can be seen that along $\rho$ direction, the spin density increases from a negative value at the center to a positive value at a certain radius, and then gradually decreases to zero.
Both the radius with the positive maximal spin density and the radius where the spin density becomes zero increase with $N$.
This feature can be understood in terms of the difference between the densities of $\uparrow$ and $\downarrow$ atoms shown in Fig.~\ref{denrz}. That the spin density is negative in inside regimes  while positive in outside regime is because we have assumed $\xi^{\alpha\alpha}_{\uparrow\uparrow} > \xi^{\alpha\alpha}_{\downarrow\downarrow}$,  consequently more  $\uparrow$ atoms tend to stay in the outside regime where the density is lower because of trapping potential, while  more  $\downarrow$  atoms tend to stay in the inside regime,  in order to lower the total energy.
Of course the spin density approaches zero for large enough value of $\rho$, as the densities of $\uparrow$ and $\downarrow$ atoms both approach zero.
Along $z$ direction with $\rho=0$, the spin density is mostly negative, also because more $\uparrow$ atoms than  $\downarrow$ atoms  stay in the larger $\rho$ regime. This effect weakens with the increase of $z$,  because the trapping potential increases, consequently the difference between the numbers of $\downarrow$ and  $\uparrow$ atoms  decreases. There is only a very small regime where the spin density becomes positive but the values are too small to be visible on the plots. This feature is unlike the that of the profiles along $\rho$ direction, as there must be some regime of $\rho$ with more $\uparrow$  atoms.

\subsection{The case with interspecies spin exchange}

Now we come to the effect of interspecies spin exchange, i.e. the case of $\xi_e\neq 0$.    We have chosen $\xi_e$=0.53 nm, 1.07 nm, 2.03 nm, 4.27 nm. As stated in the Introduction, the mean-field spin-exchange interaction energy
\begin{equation}
-2g_e \psi_{a\uparrow} \psi_{a\downarrow} \psi_{b\downarrow} \psi_{b\uparrow }  \label{at}
\end{equation}
is an interference term and acts like an attractive interaction among the four orbital wave functions in some way, but it depends on the wave functions rather than the densities. This interference effect is manifested in the number and spin density profiles.

\begin{figure*}
\scalebox{0.75}[0.75]{\includegraphics{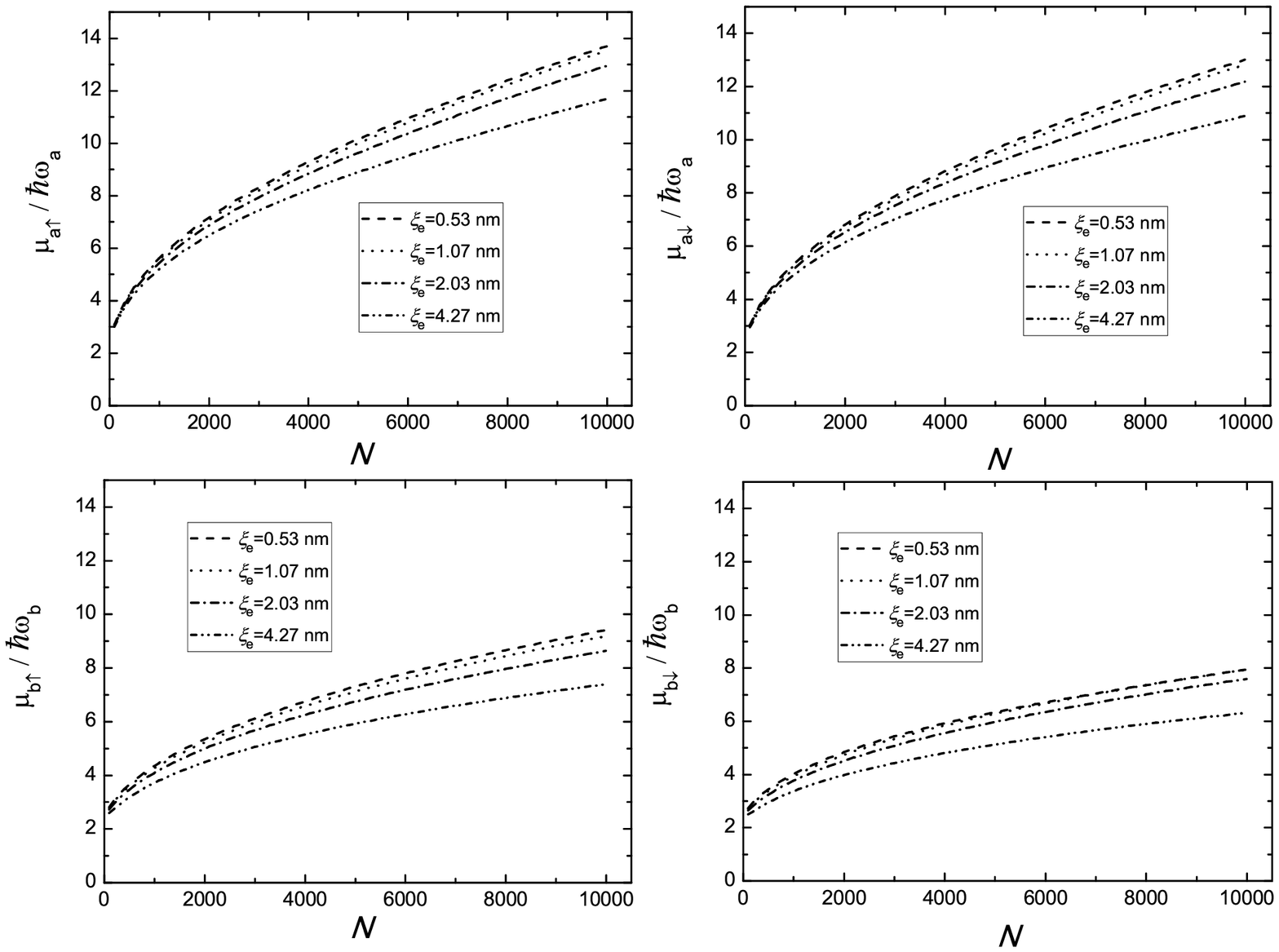}}
\caption{Reduced chemical potential $\mu_{\alpha\sigma}/\hbar\omega_{\alpha}$ varying with $N$  for four different values of $\xi_e$. The chemical potential is lowered by the spin-exchange energy, which is attractive. $d\mu_{\alpha\sigma}/d N$ remains positive.} \label{uNe}
\end{figure*}

The spin exchange lowers the chemical potential, as evident in Fig.~\ref{uNe}, which is the numerical result of $\mu_{\alpha\sigma}$ as a function of $N$ for  four values of $\xi_e$.  It can be seen from the plots that the larger  $\xi_e$ is, the lower $\mu_{\alpha\sigma}$ is. However, in all these cases,    $d\mu_{\alpha\sigma}/d N$ remains positive.

\begin{figure*}
\scalebox{0.75}[0.75]{\includegraphics{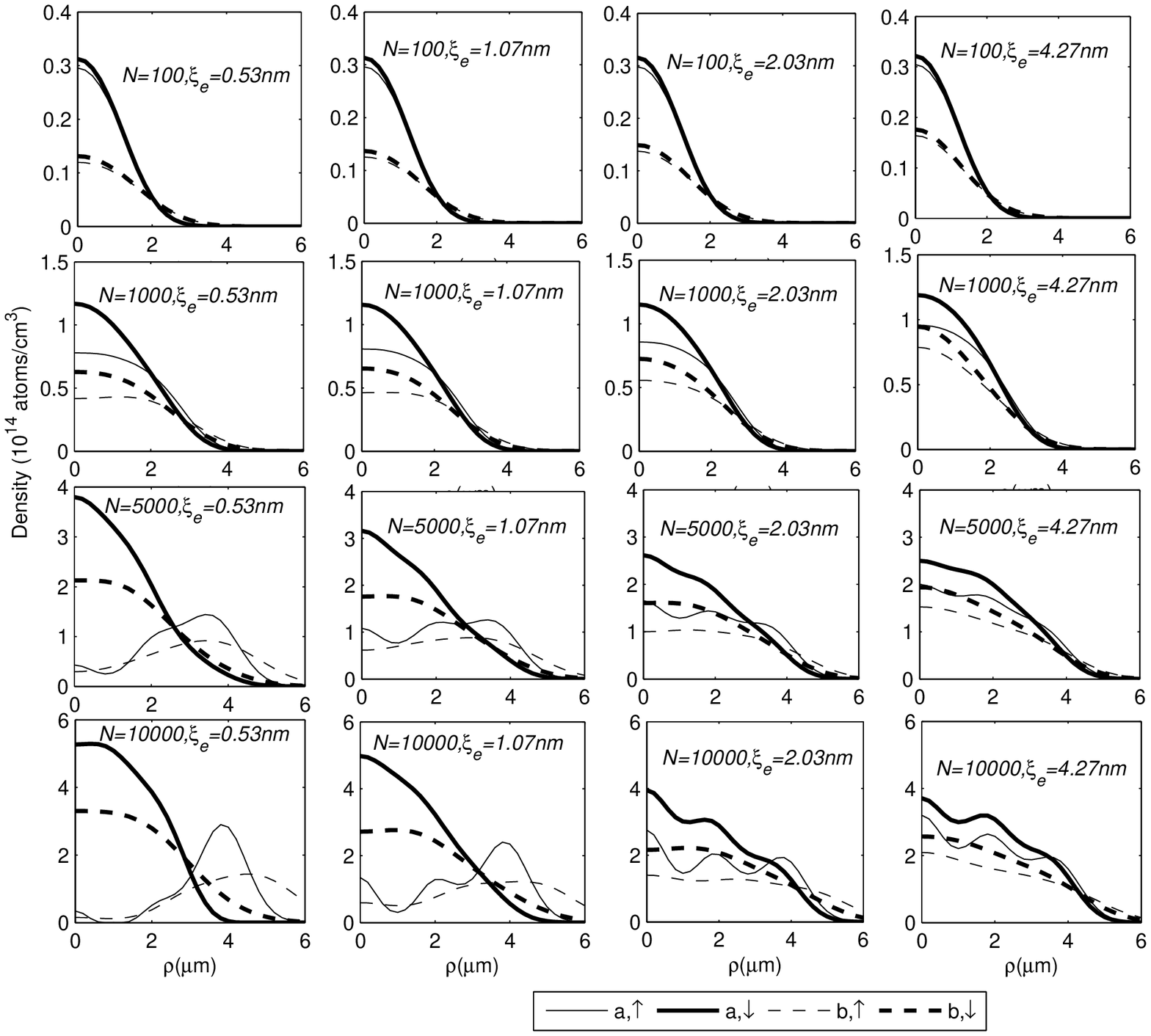}}
\caption{Density profile  $n_{\alpha\sigma}=|\psi_{\alpha\sigma}|^2$ for each pseudospin component $\sigma$ of each species $\alpha$  along $\rho$ direction on the plane $z=0$ for  $N=100, 1000, 5000, 10000$ and several values of $\xi_e$. Other parameter values are given in the main text. The larger $\xi_e$, the stronger overlap among the four wave functions. } \label{denre}
\end{figure*}

\begin{figure*}
\scalebox{0.75}[0.75]{\includegraphics{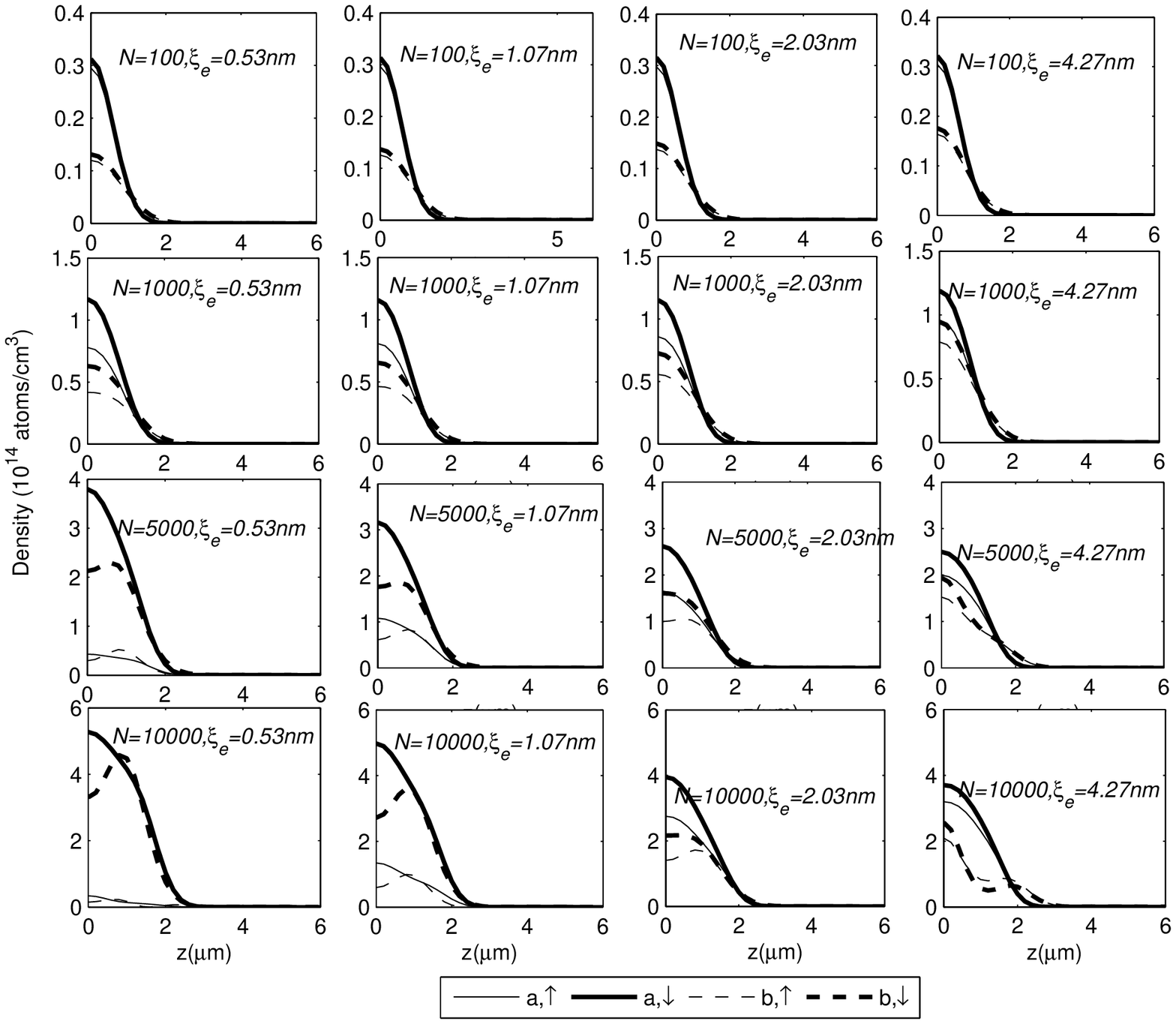}}
\caption{Density profile  $n_{\alpha\sigma}=|\psi_{\alpha\sigma}|^2$ for each pseudospin component $\sigma$ of each species $\alpha$  along $z$ direction on the line $\rho=0$ for  $N=100, 1000, 5000, 10000$ and several values of $\xi_e$. Other parameter values are given in the main text. The larger $\xi_e$, the stronger overlap among the four wave functions. } \label{denze}
\end{figure*}

\begin{figure*}
\scalebox{0.75}[0.75]{\includegraphics{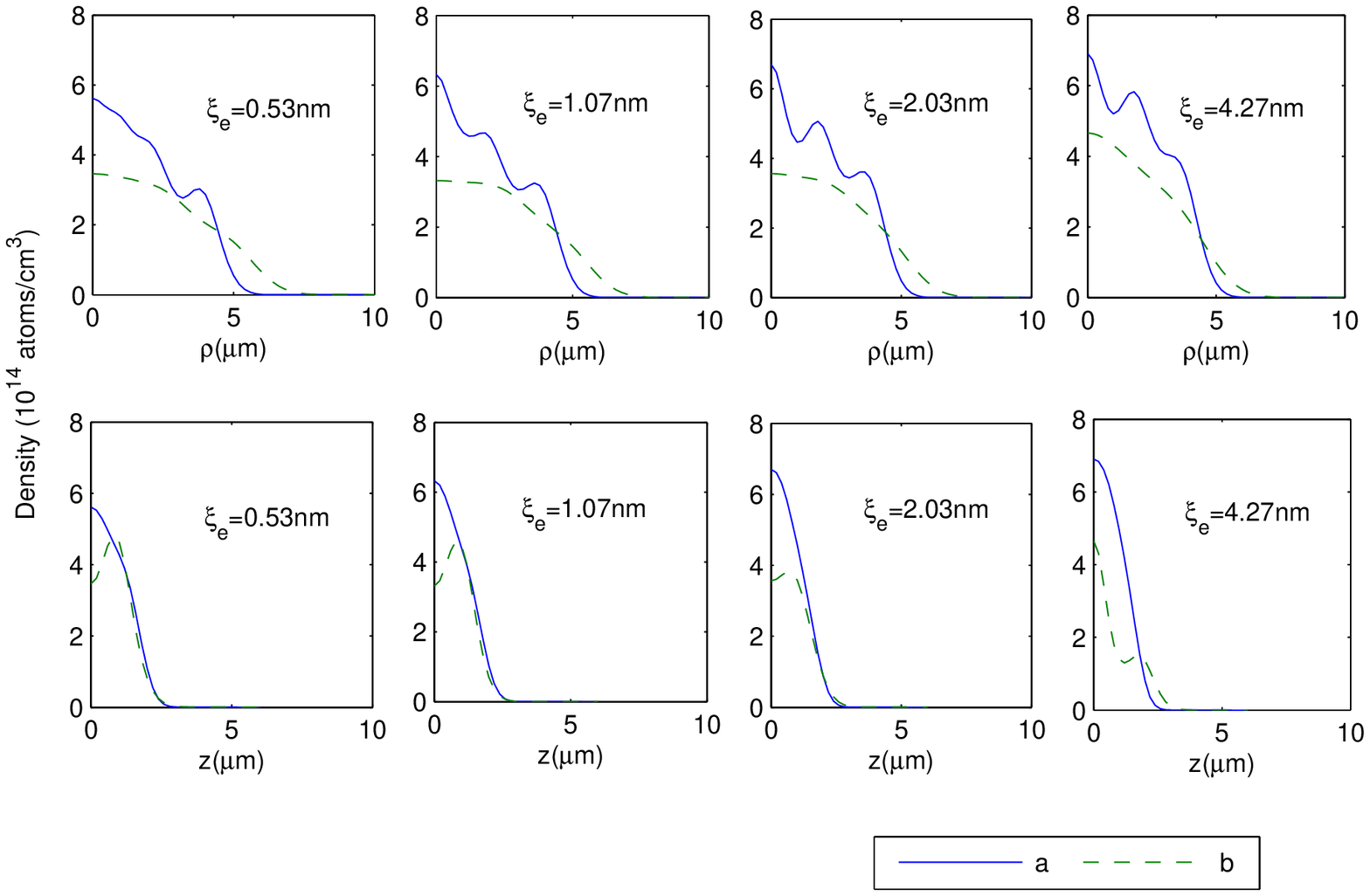}}
\caption{Profiles of the total density $n_{\alpha}$ for each species $\alpha$  along $\rho$ direction on the plane $z=0$ and on the line $\rho=0$  for atom number $N=10000$ and several values of $\xi_e$.  The larger $\xi_e$, the closer the profiles of the two species. } \label{denrze01}
\end{figure*}

The interference among the four orbital wave functions is also manifested in the  density profiles, as shown in Fig.~\ref{denre}, Fig.~\ref{denze}  and Fig.~\ref{denrze01}. The spin exchange energy  (\ref{at}) is minimized when $\psi_{\alpha\uparrow} = \psi_{\alpha\downarrow}$ for each species $\alpha$.  It  is  competitive with the other interactions. If it dominates the energy, in order to lower the energy, the wave functions of the two pseudospin components of each species  tend to be close to each other, compared with the case without spin exchange. Indeed, this tendency can be observed in  Fig.~\ref{denre}   and Fig.~\ref{denze}, which show the density profiles for each pseudospin component of each species, in $\rho$ direction with $z=0$ and in $z$ direction with $\rho =0$ respectively.  The larger  $\xi_e$ is, the more dominant the spin-exchange interaction is, then the closer  $\psi_{\alpha\uparrow}$ and $\psi_{\alpha\downarrow}$ are to each other. When  $\xi_e$ is large enough, the overlapping effect becomes very visible. Nevertheless, the overlap is not complete,  because of other terms in the energy.

Fig.~\ref{denrze01} depicts the profile of the total density of each species, which can be compared with the plot for the same $N$ in Fig.~\ref{denrz01}, it can be seen that the overlap regime of the two species is also enhanced by spin exchange,   because of the effect of spin exchange term (\ref{at}).

\begin{figure*}
\scalebox{0.75}[0.75]{\includegraphics{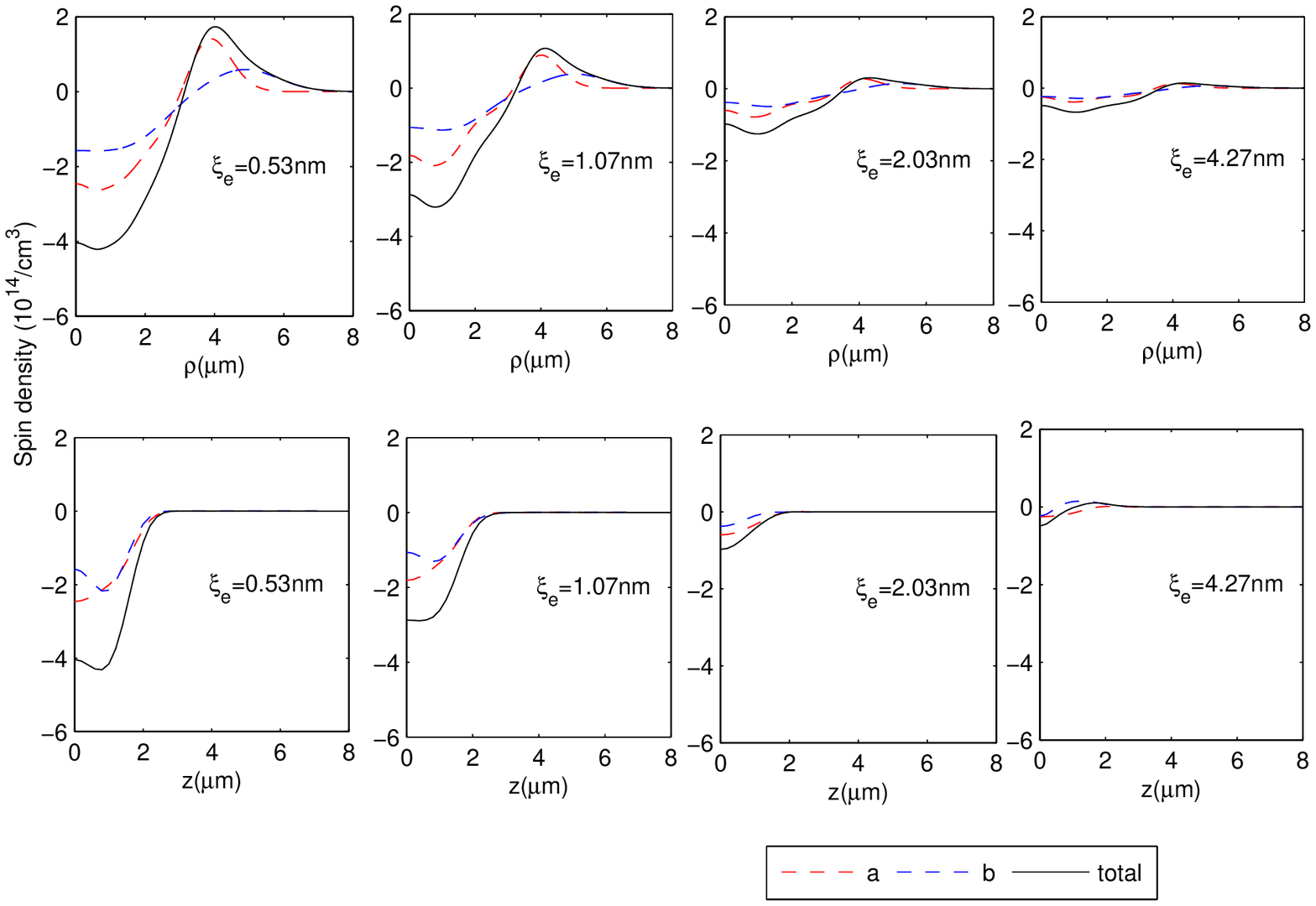}} ги
\caption{Spin density profiles in a generic parameter point as defined in the main text, for $N=10000$ and  various values of  $\xi_e$.  The larger $\xi_e$ is, the close the spin density is to $0$, even at relatively short distances from the origin. } \label{spinrze}
\end{figure*}

Moreover, we have also calculated the spin density profile, as shown in
Fig.~\ref{spinrze} for $N=10000$.  Spin density  ${\cal S}_{\alpha z} = \frac{1}{2}(|\psi_{\alpha\uparrow}| -|\psi_{\alpha\downarrow}|^2)$ is proportional to the difference between the densities of the two pseudospin components, so it is a quantification of the overlap between the two pseudospin components. Evidently, with the increase of $\xi_e$, the variation of the spin density with the radius decreases. In other words, both spin density ${\cal S}_{z\alpha}$ of each species  and the total spin density  ${\cal S}_z$ are homogenized by the spin exchange. When $\xi_e$ is large enough, each  spin density tends to vanish.
The underlying reason is also because  $\psi_{\alpha\uparrow}$ and $\psi_{\alpha\downarrow}$, whose difference gives the spin density of species $\alpha$, tend to be close to each other in order to lower the spin-exchange energy. Consequently the spin density of each species $\alpha$  tend to vanish. In both the case without spin exchange and the case with spin exchange, the location on $z$-axis where the spin density becomes zero is much smaller than that along $\rho$-direction.
Compared with the case without spin exchange, another notable feature is that when $\xi_e$ and $N$ are large enough,  in the spin density profile along $z$ direction with $\rho =0$, the regime with positive spin density becomes more visible (Fig.~\ref{spinrze}).  This is because the negative sign of the spin-exchange interaction counteracts the trapping potential, even though the trapping is stronger in $z$ direction than in $\rho$ direction.

\section{summary  \label{summary} }

This paper concerns the ground state properties of a mixture of two species of pseudospin-$\frac{1}{2}$ Bose gases with interspecies spin-exchange interaction in a trapping potential. We have numerically calculated the four orbital condensate wave functions, each of which corresponding to  a pseudospin component of each species, by using the  GP-like equations. We set the atom number of each species to be $N$. Using these wave functions, the number and spin densities are obtained.
When the spin-exchange scattering length is zero, this mixture reduces to a mixture of the usual type, with  four components. Various features appear as consequences of minimizing the density-density interaction. For example, with the increase of $N$, the difference between $n_{\alpha\uparrow}$ and $n_{\alpha\downarrow}$ for each species $\alpha$ increases.

If there exists  interspecies spin-exchange scattering,
novel features absent in  the usual mixtures emerge. As the spin-exchange interaction is negative as a consequence of minimizing the energy, it acts like an attractive interaction. Nevertheless, it depends on the overlap among the four wave functions.   It  lowers the chemical potentials and make the densities of the two pseudospin components of each species  tend to be close to each other, and thus the spin density tends to be homogenized, and even tends to vanish when the spin-exchange scattering length is so large that it dominates over the density-density interaction.

Therefore as experimentally measurable quantities, the number and spin density profiles of such a mixture with interspecies spin exchange are effective probes of  the novel many-body ground state of this system.

\begin{acknowledgments}

We thank Li Ge  for useful discussion.
This work was supported by the National Science Foundation of China (Grant No. 11074048)  and the Ministry of Science and Technology of China (Grant No. 2009CB929204).

\end{acknowledgments}

\end{document}